\documentclass[aps,prd,twocolumn,notitlepage,10pt,
nofootinbib,nobibnotes,superscriptaddress,preprintnumbers]{revtex4-1}

\usepackage{amsmath}
\usepackage{amssymb}
\usepackage{graphicx,xcolor} 
\usepackage[normalem]{ulem}
\usepackage{appendix}

\begin{document}
\title{Expansion attractor in classical Yang-Mills theory}
\author{Clemens Werthmann}
\affiliation{Department of Physics and Astronomy, Ghent University, 9000 Ghent, Belgium}

\date{\today}

\begin{abstract}
    Applications of the concept of a hydrodynamic attractor have been mostly constrained to the context of hydrodynamization in heavy ion collisions. Deepening the understanding of the general phenomenon requires generalizing it to other contexts in order to see which facettes are truly universal and not a peculiarity of equilibrating conformal Bjorken flow. As part of this endeavor, this work aims to study the behaviour in a system that does not hydrodynamize, but nevertheless loses memory through expansion. We find that the system goes through three stages. At early times, expansion drives the ratio $P_L/\epsilon$ to converge to $-1$ as $\tau^{-2}$. This memory loss is partially restored at intermediate times, when interaction dominated degrees of freedom through expansion dominated evolution grow quadraticallly in time. At late times, interactions shuffle of information in state space without significant memory loss or restoration.
\end{abstract}

\maketitle

\section{Introduction}

The paradigm of hydrodynamic attractors~\cite{Heller:2015dha} was introduced in an effort to explain the success of hydrodynamic descriptions of heavy ion collisions, on time and length scales where the required conditions for its theoretical construction may not be fulfilled. The idea is that long before local equilibration, the system already exhibits a universal time evolution across a wide range of initial conditions. In conformal Bjorken flow, this fact is apparent in the time evolution of many individual quantities, for example the pressure anisotropy, quickly converging to a universal curve. 

Indeed, descriptions of Bjorken flow in Mueller-Israel-Steward-type hydrodynamics~\cite{Heller:2015dha,Romatschke:2017acs}, AdS/CFT~\cite{Spalinski:2017mel} and kinetic theory~\cite{Heller:2016rtz,Boguslavski:2023jvg} all exhibit this behaviour. All of these are dynamical descriptions that are viable in simulations of heavy ion collisions. In many systems, these attractor curves can be understood as a dynamical interpolation of two fixed points~\cite{Blaizot:2020gql,Blaizot:2021cdv}. At late times, interactions draw the system towards local equilibrium, while at early times, it is the strong longitudinal expansion that drives the system to a specific state. This longitudinal expansion usually does not affect details of the transverse degrees of freedom, such as the distribution of transverse momenta in kinetic theory, such that the early stage will cause only partial memory loss. Equilibration on the other hand facilitates memory loss in all of state space, except for conserved quantities, meaning that memory that has been lost in the early stages remains lost. These facts are also apparent for example in the evolution of eigenstate coefficients in the picture of adiabatic hydrodynamization~\cite{Rajagopal:2024lou,Rajagopal:2025nca,DeLescluze:2025gaa}. Part of the motivation of this paper is to characterize attractor behaviour for an example of a system where memory may be partially restored.

Most of the studies of attractor behaviour have been restricted to hydrodynamizing dynamical descriptions of heavy ion collisions. However, a more thorough understanding of the phenomenon requires extending these studies to other systems, where the attractor may take a different form. An early example was the study of attractor behaviour in Gubser flow~\cite{Denicol:2018pak,Behtash:2019qtk,Dash:2020zqx}, which transitions from longitudinal expansion at early times to transverse expansion at late times. In this setup, a third fixed point emerges at very late times, which is dominated by the transverse expansion. Systems at low energy density may not equilibrate sufficiently and skip the ideal hydrodynamic fixed point. A more recent idea is the extension of the concept to cold atom systems~\cite{Fujii:2024yce,Mazeliauskas:2025jyi}, which differ in many ways from heavy ion collisions but exhibit many of the same phenomena. More importantly, in this system, the behaviour would be directly accessible in experiment. However, a study including realistic transport coefficients~\cite{Heller:2025yxm} showed that in this case, a nontrivial coupling of the evolution of the non-equilibrium and the equilibrium sector obscure the dynamics: it is no longer apparent in the convergent evolution of individual observables, but requires tracking dimensional reduction in state space, a diagnostic that was previously introduced for Bjorken flow~\cite{Heller:2020anv}.

This work aims at further generalizing the range of dynamical descriptions in which the behaviour is studied. Another such description that is relevant to heavy ion collisions is the one of strong fields in the earliest stages. At high collision energies, the makeup of nuclei is dominated by gluons, which are so high in number that their evolution can be described via classical Yang-Mills fields, while quarks become static sources moving along the light cone~\cite{McLerran:1993ni,McLerran:1993ka,Gelis:2012ri,Schlichting:2019abc,Garcia-Montero:2025hys}. Quantum fluctuations are taken into account by considering an independently evolving ensemble of fluctuating initial conditions. Just after the collision, the system may be further evolved via Yang-Mills dynamics in an evolutionary stage known as the Glasma~\cite{Kovner:1995ts,Kovner:1995ja,Lappi:2006fp}. While with a nonthermal fixed point, these dynamics possess a dynamically attractive structure, this description does not hydrodynamize~\cite{Schlichting:2012es,Kurkela:2012hp}, and a recent study~\cite{Pooja:2026hjg} has found, much in the spirit of what will be discussed in this work, that after an initial period of decay, it exhibits chaotic behaviour and exponential growth of differences between neighbouring trajectories. We will discuss 

While motivated by it, the dynamics that will be studied in this work are not exactly those of the theory described above, as we are searching for a simple example case of a system that exhibits qualitatively new attractor behaviour. Rather than studying an ensemble of fluctuating initial conditions which evolve independently according to Yang-Mills equations, we will go the simpler route of studying fully classical Yang-Mills theory with no fluctuations. In the case of gauge-invariant descriptions, the color fields must have a zero expectation value, $\langle A_a^\mu\rangle=0$, which Elitzur's theorem states for the case of lattice gauge theory. We will work in a gauge fixed description, where nonzero expectation values are allowed.

The case considered here amounts to a system that is dominated not by particle-like excitations but by strong coherent fields, like in Lund strings~\cite{Andersson:1983ia,Lappi:2006fp}. Thus, it can be viewed as something akin to the early stage of the central region of a flux tube generated in a $e^+e^-\to q\bar{q}$ collisions, though in the present case, we do not assume transverse isotropy. We do impose, however, all other Bjorken-like symmetries and then search for and study attractor behaviour. 

The fact that we consider a non-fluctuating system means that the known attraction to a non-thermal fixed point~\cite{Berges:2013eia,Berges:2013fga} is not present in this system. Also, in the expansion-less case, it is known that Yang-Mills theory exhibits linear instabilities known as Nielsen-Olesen instabilities~\cite{Nielsen:1978rm,Matinyan:1981dj,Chirikov:1981cm}, characterized by exponential growth of differences between neighbouring curves, which would compete with a potential memory loss mechanism. This has also been studied more recently for the case of a background that is uniform in space -- much like in this paper -- but also uniform in time, both for the flat~\cite{Bazak:2021xay} and the expanding case~\cite{Bazak:2023kol}. However, the unstable perturbations do have a structure in space. Since we consider transverse homogeneity, we do not expect these instabilities to occur here. We note however, that this means that we consider an idealized model system that would immediately collapse in reality. The behaviour that we do expect is expansion driven memory loss, and as it turns out the system studied here also exhibits a partial recovery of memory in a predictable way. This makes it an interesting model system for the general study of attractor behaviour. We will work with $N_c=2$, which not only simplifies the problem but is in fact necessary for part of the setup. 

This paper is structured as follows. Section~\ref{sec:setup} introduces the system that is studied in this work, including symmetry assumptions, evolution equations and constraints. The equations allow for analytical solutions at early times, which seem to exhibit expansion driven attractor behaviour as discussed in Sec.~\ref{sec:early_times}. Section~\ref{sec:glasma} then discusses similarities in the scaling and differences in the early times to the Glasma description. In Sec.~\ref{sec:pressure_anisotropy} we discuss the phenomenology of the pressure anisotropy and how the attractor behaviour manifests in this observable. However, the full state of the system covers many more degrees of freedom, so a full discussion of attractor behaviour requires tracking dimensional reduction in state space as discussed in Sec.~\ref{sec:dimensional_reduction}. Sec.~\ref{sec:late_time} deals with the behaviour at very late times and discusses to what extent the observed exponential growth is present also in this system. We conclude with Section~\ref{sec:conclusions}.

\section{Setup}\label{sec:setup}

The degrees of freedom of classical Yang-Mills theory are the classical non-abelian gauge fields $A^\mu=A_a^\mu T^a$, with $a$ being the color index and $T^a$ being the generators of the gauge group. Their commutation relations are given by the structure constants $f^{ab}_c$: $[T^a,T^b]=if^{ab}_cT^c$. We consider the sourceless case, so the evolution is given in terms of the field strength tensor
\begin{align}
    F^{\mu\nu}=\nabla^\mu A^\nu-\nabla^\nu A^\mu-ig[A^\mu,A^\nu]
\end{align}
as
\begin{align}
    D_\mu F^{\mu\nu}=0\,,\label{eq:evolution_basic}
\end{align}
where we defined $D_\mu(\cdot)=\nabla_\mu(\cdot)-ig[A_\mu,\cdot]$. We consider Milne coordinates $(\tau,x,y,\eta)$, which are adapted to boost invariance. The proper time $\tau=\sqrt{t^2-z^2}$ is invariant and the spacetime rapidity $\eta=\mathrm{artanh}(z/t)$ is additive under boosts in the z-direction. In these coordinates, $g_{\mu\nu}=\mathrm{diag}(1,-1,-1,-\tau^2)$.

We enforce Bjorken-like symmetries by requiring that the fields $A^\mu$ are boost-invariant and homogeneous in the transverse plane, i.e. $A^\mu(\tau,x,y,\eta)=A^\mu(\tau)$. This means that all gradients in $x$, $y$ and $\eta$ vanish. Full Bjorken symmetries would also require isotropy in the transverse plane. When enforcing this on the level of the energy-momentum-tensor, $T^{xx}=T^{yy}$ and $T^{xy}=0$, the resulting conditions are not conserved by the evolution equations. On the other hand, enforcing it on the level of $F_a^{\mu\nu}$ requires that all $F_a^{\mu i}$ vanish. This, however, would immediately render the system static. We therefore choose to consider the nontrivial case of an absence of transverse isotropy. Together with the gauge fixing condition $A^\tau=0$, the equations~\eqref{eq:evolution_basic} become
\begin{align}
    0&=f_a^{bc}(A_b^i\partial_\tau A_c^i+\tau^2A_b^\eta\partial_\tau A_c^\eta)=C_a\,,\label{eq:evolution_t}\\
    0&=\partial_\tau^2A_a^i+\tau^{-1}\partial_\tau A_a^i-g^2f_a^{bc}f_c^{de}(A^{\bar{i}}_bA^{\bar{i}}_d+\tau^2A_b^\eta A_d^\eta)A_e^i\,,\label{eq:evolution_i}\\    
    0&=\partial_\tau^2A_a^\eta+3\tau^{-1}\partial_\tau A_a^\eta -g^2 f_a^{bc}f_c^{de}A^i_bA^i_dA_e^\eta\,,\label{eq:evolution_eta}
\end{align}
where here and in the following, $i,j\in\{x,y\}$, $u,v\in\{x,y,\eta\}$ and $\bar{i}=x$ for $i=y$ and vice versa. By the structure of these equations, we can implicitly non-dimensionalize the problem by viewing $\tau\to\tau_{\rm scale}\tau$, $A_a^i\to\tau_{\rm scale}^{-1}A_a^i$ and $A_a^\eta\to\tau_{\rm scale}^{-2}A_a^\eta$. Similarly, the coupling $g$ can be absorbed into the fields $A_a^u$, showing that in terms of the time evolution it has the same effect as changing the scale of the initial conditions. This reparametrization would, however, change how physical quantities are expressed in terms of the fields $A_a^u$, which still depends on g. We choose to keep the factor of $g$ as-is in order to more easily identify terms related to interaction. For numerical results, we pick $\alpha_s=\frac{g^2}{4\pi}=0.3$.

Now, the above are four coupled mixed order differential equations for three remaining unknown fields $A^i$, $A^\eta$. However, Eqs.~\eqref{eq:evolution_i} and~\eqref{eq:evolution_eta} guarantee that the derivative of $C_a$ in Eq.~\eqref{eq:evolution_t} fulfills $\partial_\tau C_a=-C_a/\tau$, meaning that if $C_a=0$ at initial time $\tau_0$, then it remains so at all times. Thus, Eq.~\eqref{eq:evolution_t} is actually a constraint on the initial conditions. What remains is a system of three second order differential equations for three unknowns.

While the $A^\mu$ are the underlying degrees of freedom of the system, it is convenient to specify the state of the system in terms of the elements $F^{\mu\nu}_a$ of the field strength tensor. A set of manifestly gauge invariant observables could be constructed as the possible colour scalar products of the components of $F^{\mu\nu}_a$. However, some of the behaviour we will discuss is more directly apparent in the components themselves, so we choose to consider them as our observables.   They relate to the components of the energy-momentum tensor as
\begin{align}
    T^{\mu\nu}&=\mathrm{tr}\left(F^{\mu\rho} F^{\nu}_\rho-\frac{1}{4}g^{\mu\nu} F^{\rho\sigma}F_{\rho\sigma}\right)\,.
\end{align}
As a consequence, defining  transverse and longitudinal components of the chromoelectric and -magnetic fields as
\begin{align}
    E_a^i&=F^{\tau i}_a\,, \quad &E_{L,a}=\tau F^{\tau\eta}_a\,, \label{eq:Edef}\\ B^i_a&=\tau F^{\bar{i}\eta}_a\,, \quad &B_{L,a}=F^{xy}\,,\label{eq:Bdef}
\end{align}
 we can write the pressures and energy density as
\begin{align}
    e&=\frac{1}{2}\left(E_{L,a}E_L^a+B_{L,a}B_L^a+\sum_iE_{a}^iE^{i,a}+B_a^iB^{i,a}\right)\,,\\
    P_L&=\frac{1}{2}\left(-E_{L,a}E_L^a-B_{L,a}B_L^a+\sum_iE_{a}^iE^{i,a}+B_a^iB^{i,a}\right)\,,\\
    P_i&=P_T+\frac{1}{2}\left[E_{a}^{\bar{i}}E^{\bar{i},a}+B_a^{\bar{i}}B^{\bar{i},a}-E_{a}^iE^{i,a}-B_a^iB^{i,a}\right]\,,\\
    P_T&=\frac{1}{2}\left(E_{L,a}E_L^a+B_{L,a}B_L^a\right)\,.
\end{align}
The last line defines the average transverse pressure.

Under the enforced symmetries and with the axial gauge choice $A^\tau=0$, the components of $F^{\mu\nu}_a$ are given as
\begin{align}
    F^{\tau i}_a&=\partial_\tau A_a^i\,,\label{eq:Ftaui}\\
    F_a^{\tau\eta}&=\partial_\tau A_a^\eta+2\tau^{-1}A_a^\eta\,,\label{eq:Ftaueta}\\
    F_a^{uv}&=gf_a^{bc}A_b^uA_c^v\,.\label{eq:Fuv}
\end{align}
Note that in the present case, electric fields only have abelian components, while magnetic fields only have non-abelian components and are proportional to the coupling $g$. Since we use these fields as coordinates in the state space of this theory, we also have to invert these relations in order to translate initial conditions. For known $A_a^\eta$, Eqs.~\eqref{eq:Ftaui} and \eqref{eq:Ftaueta} can immediately be solved for the $\partial_\tau A_a^u$, but Eq.~\eqref{eq:Fuv} is not easily inverted. As discussed in Appendix~\ref{app:inverting}, the inversion is possible only in the case of $SU(2)$. This is another reason why we consider the case $N_c=2$ here. Specifically, the inverted relations are

\begin{align}
    A_a^u&=\frac{1}{\sqrt{N}}f_a^{bc}F_b^{u,(u+1)\%3}F_c^{(u-1)\%3,u}\,,\\
    N&=gf^{abc}F_a^{xy}F_b^{y\eta}F_c^{\eta x}\,.
\end{align}

Finally, the system still has a remaining global gauge freedom of $A\to UAU^\dagger$ for any constant $U\in SU(2)$. Unlike $A^\tau = 0$, this residual freedom cannot be fixed by a condition that the evolution preserves. Such a condition would have to align some conserved adjoint-valued quantity with a fixed color direction. The only candidate is the total color charge, which is the Noether charge of precisely this residual symmetry. But that charge is Eq.~\eqref{eq:evolution_t}, which vanishes on physical configurations. The symmetry one wishes to fix therefore supplies no invariant color frame to fix it against, and any gauge condition on the $A^u_a$ is necessarily an algebraic slice imposed at a single time. Thus, we impose only at initial time the additional constraint 
\begin{align}
    A^x_2 = A^x_3 = A^y_3 = 0\,,\ A^x_1 > 0\,,\ A^y_2 > 0\,,\label{eq:gaugeA}
\end{align}
which is equivalent to
\begin{align}
    B_{L,1} = B_{L,2} = 0, B^y_1 = 0\,, B_{L,3} > 0\,, B^y_2 > 0\,.\label{eq:gaugeB}
\end{align}
Note that these will not remain true at later times. However, again since Eq.~\eqref{eq:evolution_t} is the Noether charge of global color rotations, the gauge degrees of freedom are tangential to this constraint, such that we can always project them out also at later times. In total, at any time, the space of physical configurations we consider is $12$-dimensional, but the way this subspace is embedded in the $18$-dimensional state space that remains after imposing the axial gauge is nontrivial and changes over time, so we track all $18$ variables.

\section{Early Times}\label{sec:early_times}

We can try to anticipate the early time behaviour by considering the case $\tau\ll1$, which amounts to neglecting the terms containing no factor of $\tau^{-1}$ in Eqs.~\eqref{eq:evolution_i} and~\eqref{eq:evolution_eta}, i.e. the interaction terms. For this case, analytical solutions can be obtained even in the Glasma case with transverse structure by working in Fourier space~\cite{Epelbaum:2013waa,Berges:2013fga}. The reason that this case is much simpler is that it amounts to considering free fields that each evolve on their own, i.e. the equations decouple. In our case, the solutions are:

\begin{align}
    A_a^i(\tau)&=A_a^i(\tau_0)-\tau_0\partial_\tau A_a^i(\tau_0)\ln\left(\frac{\tau}{\tau_0}\right)\,,\\
    A_a^\eta(\tau)&=A_a^\eta(\tau_0)-\frac{\tau_0}{2}\partial_\tau A_a^\eta(\tau_0)\left[\left(\frac{\tau}{\tau_0}\right)^{-2}-1\right]\,.
\end{align}

Despite this simplification, the phenomenology is more complex than in the corresponding cases of MIS hydro or kinetic theory, because the chromoelectric and -magnetic fields are combinations of these solutions, and the pressures are again squares of these. Specifically, the fields are given as

\begin{align}
    E_a^i&=\partial_\tau A_a^i(\tau_0)\frac{\tau_0}{\tau}\,,\label{eq:ana_ET}\\
    E_{L,a}&=2A_a^\eta(\tau_0)+\tau_0\partial_\tau A_a^\eta(\tau_0)\,,\label{eq:ana_EL}\\
    B_a^i&=g\left[A_a^{\bar{i}}(\tau_0)-\tau_0\partial_\tau A_a^{\bar{i}}(\tau_0)\ln\left(\frac{\tau}{\tau_0}\right)\right]\nonumber\\
    &\times\left\{\tau A_a^\eta(\tau_0)-\frac{1}{2}\partial_\tau A_a^\eta(\tau_0)\left[\frac{\tau_0^3}{\tau}-\tau_0\tau\right]\right\}\,\label{eq:ana_BT}\\
    B_{L,a}&=g\left[A_a^x(\tau_0)-\tau_0\partial_\tau A_a^x(\tau_0)\ln\left(\frac{\tau}{\tau_0}\right)\right]\nonumber\\
    &\times\left[A_a^y(\tau_0)-\tau_0\partial_\tau A_a^y(\tau_0)\ln\left(\frac{\tau}{\tau_0}\right)\right]\,.\label{eq:ana_BL}
\end{align}

They display a wide range in phenomenology. Transverse electric fields decay as $\tau^{-1}$ while longitudinal electric fields stay constant. Depending on the initial values, transverse magnetic fields transition from a decay $\propto\tau^{-1}$ to a linear increase in $\tau$, which comes both with and without an additional factor $\ln(\tau/\tau_0)$. Longitudinal magnetic fields contain constant terms as well as terms that are linear and quadratic in $\ln(\tau/\tau_0)$. The interplay of decay, growth and stagnation will help us with the interpretation of results in the following.

Let us now discuss what this early time behaviour means for the pressure anisotropy. For this, we simplify the time dependence by neglecting any terms that come with any factor of $\tau_0$ that can not be compensated by a factor of $\tau^{-1}$. Note that this means we consider the generic case, while in some special cases the choice of initial conditions may render the otherwise leading terms subleading. We find that longitudinal fields are constant, while transverse electric fields evolve $\propto\tau_0/\tau$. In terms of this power counting, one should view $A_a^\eta(\tau_0)\sim\tau_0^{-1}$ because of the metric factor $g_{\eta\eta}=-\tau^2$. Thus, transverse magnetic fields carry a factor of $\tau/\tau_0$. 

Now writing
\begin{align}
    \frac{P_L}{\epsilon}=-1+\frac{2E_T^2+2B_T^2}{E_L^2+B_L^2+E_T^2+B_T^2}\,,
\end{align}
by the time dependence of the different field components, we can identify four regimes:
\begin{align}
    E_T^2\gg E_L^2+B_L^2\gg B_T^2: &\frac{P_L}{\epsilon}\approx 1\,,\\
    E_L^2+B_L^2\gg E_T^2\gg B_T^2: &\frac{P_L}{\epsilon}\sim -1+k_1\tau_0^2/\tau^2\,,\label{eq:regime_decay}\\
    E_L^2+B_L^2\gg B_T^2\gg E_T^2: &\frac{P_L}{\epsilon}\sim -1+k_2\tau^2/\tau_0^2\,,\label{eq:regime_growth}\\
    B_T^2\gg E_L^2+B_L^2\gg E_T^2: &\frac{P_L}{\epsilon}\approx 1\,.
\end{align}
Note that $B_T^2\propto g^2$ and therefore the transition from the second to the third regime occurs earlier for stronger coupling. Thus, despite not yet playing a role in the dynamics of individual fields, interactions are responsible for the transition from a power law decrease to a power law growth of $P_L/\epsilon$. The transition occurs when $E_T^2\approx B_T^2$. Using Eqs.~\eqref{eq:ana_ET} and~\eqref{eq:ana_BT} with $\tau_0\approx 0$, we find for the transition timescale
\begin{align}
    \tau_{\rm trans} \approx \left(\frac{\tau_0^2\sum_i\partial_\tau A_a^i\partial_\tau A^{i,a}(\tau_0)}{g^2\sum_if_a^{bc}A_b^iA^{\eta}_cf^{ade}A_d^iA^{\eta}_e(\tau_0)}\right)^{1/4}\,.\label{eq:tau_trans}
\end{align}

\section{Making contact with the Glasma}\label{sec:glasma}

In the Glasma, the fields are sourced by the static sources in the colliding nuclei. From the nuclei, there is one dimensionful scale that enters and determines parametric dependencies: the nuclear saturation scale $Q_s$~\cite{Schlichting:2019abc}. The fluctuating colour densities in the nuclei are usually modeled to be Gaussian and have a variance of $\langle\rho \rho\rangle\sim g^2\mu^2$~\cite{Kovner:1995ts}, where $\mu$ is the colour charge density per unit area, which is equal to the number of colour charges per nucleon times the nuclear thickness function and satisfies $Q_s\approx 0.6 g^2\mu$. The charges then source the colour fields, such that one may estimate $gA\sim Q_s$. Another way to arrive at this parametric relation is via the energy density~\cite{Lappi:2006hq}: $g^{-2}Q_s^4\sim e\sim (F^{\mu\nu})^2\sim (gA^2)^2$. The scale $Q_s$ also controls fluctuations in the form of the inverse correlation length of the fields.

In the system considered in this work, there is no transverse structure and therefore no correlation length, but as noted already in the discussion below Eqs.~\eqref{eq:evolution_t}--\eqref{eq:evolution_eta}, also here the dynamical timescale is dictated by the size of the fields and the coupling, and it makes sense to think about it as $Q_s\sim gA^{\tau/i}(\tau_0)\sim g\tau_0A^\eta(\tau_0)$. We will, however, keep the nondimensionalized description of the system, where $Q_s= g\tau_{\rm scale}^{-1}$. The results can be translated to a specific value of $Q_s$ by computing the corresponding $\tau_{\rm scale}$, which all quantities are stated in units of. For $Q_s=1\,$GeV, we find $\tau_{\rm scale}=1.94\,\mathrm{GeV}^{-1}=0.383\mathrm{fm}/c$.

In the boost-invariant Glasma, transverse fields must vanish at early times to ensure regularity at $\tau=0$. In an early time expansion, the first nonvanishing component appears at quadratic order in time and is sourced by transverse gradients~\cite{Fujii:2008km}. This is a clear distinction to the system considered here: by allowing transverse fields to take nonzero values very early on, we find decaying and linearly growing trends that are absent in the boost-invariant Glasma. On the other hand, transverse gradients play no role here and the associated quadratic behaviour in time is not observed. More realistic Glasma descriptions consider the non-boost-invariant case, accounting for the longitudinal structure of nuclei~\cite{Schlichting:2020wrv}. In these cases, transverse fields are no longer forced to vanish, such that these systems may also produce some of the same phenomenology observed in this work.

\section{Pressure Anisotropy}\label{sec:pressure_anisotropy}

When diagnosing attractor behaviour, the first indicator to check is the evolution of the pressure anisotropy, which we quantify via the ratio $P_L/e$. In Bjorken flow, the behaviour typically manifests in the form of a quick convergence to a universal time evolution curve, referred to as the attractor curve. In hydrodynamics and kinetic theory, the attractor curve interpolates between an early time expansion dominated fixed point far from equilibrium and a late time interaction dominated fixed point given by local equilibrium. The interpretation of the early time behaviour is simplest in kinetic theory. The fact that particles free-stream towards the spacetime rapidity $\eta$ that matches their own pseudorapidity $y$ causes longitudinal momenta in comoving coordinates to decrease. This is why the transverse pressure exhibits a decay towards $P_L=0$. In the present case, as discussed in the previous section, we find both decreasing and increasing behavior. So both convergence and divergence are possible at early times, depending on the precise timescale and the initial conditions. We cannot make any statements about the late time behaviour yet, because the structure is too complex to allow for an analytical treatment. Thus, we will look to numerical solutions of Eqs.~\eqref{eq:evolution_i} and~\eqref{eq:evolution_eta}.

Since the early time behaviour features power laws and does not usually have an intrinsic scale, while the late time behaviour is expected to set in on the timescale of interactions, when studying attractors it is common to express time in units of this interaction timescale. This is however not the same timescale as in Eq.~\eqref{eq:tau_trans}, but the timescale at which interactions become relevant for the underlying dynamics. Taking a closer look, Eqs.~\eqref{eq:evolution_i} and~\eqref{eq:evolution_eta} are of the form
\begin{align}
    0&=\partial_\tau^2A_a^u+k\tau^{-1}\partial_\tau A_a^u-M_a^eA_e^u\,,\\
    M_a^e&=g^2f_a^{bc}f_c^{de}A_b^uA_d^u\,,\label{eq:tau_int}
\end{align}
so we define the interaction timescale $\tau_{\rm int}$ as the inverse square root of the largest eigenvalue of the matrix $M$. As discussed in Sec.~\ref{sec:glasma}, in analogy to the Glasma it makes sense to think of dynamical timescales in terms of $Q_s\sim gA$. By the parametric scaling of $M\sim g^2A^2\sim Q_s^2$, we therefore find $\tau_{\rm int}\sim Q_s^{-1}$, exactly as the Glasma analogy suggests.

\begin{figure}
    \centering
    \includegraphics[width=\linewidth]{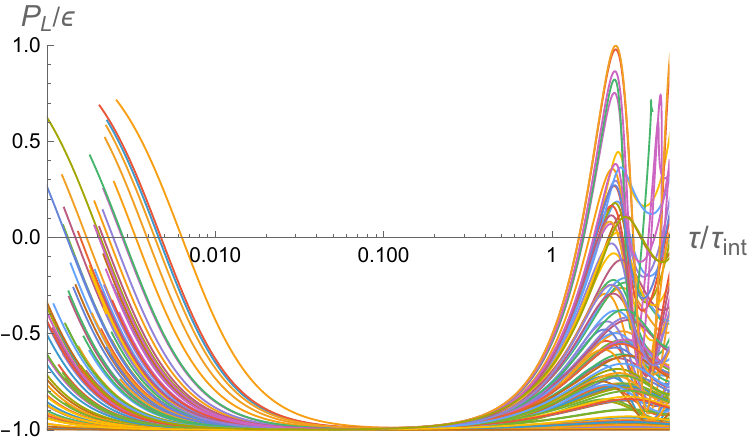}
    \includegraphics[width=\linewidth]{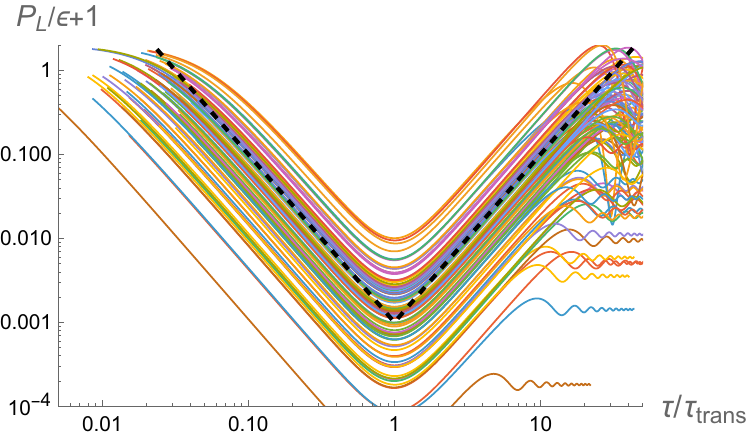}
    \caption{Early time evolution of the pressure anisotropy from numerical simulations of the dynamics prescribed by Eqs.~\eqref{eq:evolution_i} and~\eqref{eq:evolution_eta} with random initial conditions. The upper plot shows $P_L/\epsilon$ as a function of time scaled by the interaction timescale $\tau_{\rm int}$ on a logarithmic scale. The lower plot shows the deviation of $P_L/\epsilon$ from its early time attractive value $-1$ as a function of time in terms of the early power law transition timescale $\tau_{\rm trans}$ in $\log$-$\log$-scale.}
    \label{fig:pressure_early}
\end{figure}

Fig.~\ref{fig:pressure_early} shows the evolution of the pressure anisotropy, extracted from numerical solutions to the full evolution equations starting at $\tau_0=10^{-3}$. The 100 different initial conditions were obtained by random uniform sampling of values for the 12 free fields, setting 3 fields to zero as per the gauge choice, and fixing the remaining 3 fields via the conditions~\eqref{eq:evolution_t}. As argued earlier, generically one should view $A_a^\eta(\tau_0)\sim\tau_0^{-1}$. However, $A_a^\eta\gg A_a^i$ immediately means $E_L^2\gg B_T^2\gg E_T^2$, meaning the regime of~\eqref{eq:regime_decay} is skipped. Comparing with Eq.~\eqref{eq:tau_int}, it also means that $\tau_{\rm int}\sim g^{-1}\tau_0$, so interaction dominated evolution sets in very early. Thus, we pick all $A_a^{u}$ and $\partial_\tau A_a^{u}$ from the interval $[-1,1]$.

At early times, the longitudinal pressure ratio quickly drops to $P_L/\epsilon=-1$ for all initial conditions and then rises again. At late times, oscillatory behaviour sets in. In the upper plot of Fig.~\ref{fig:pressure_early}, we can see that when measuring time in terms of $\tau_{\rm int}$, the first maxima of this oscillatory behaviour align between different curves, so we can identify this as the regime where interactions dominate the dynamics. As seen in the lower plot, measuring time in terms of $\tau_{\rm trans}$ allows to align the transition from quadratic decay to quadratic growth. In its doubly logarithmic scale, the two power laws are apparent.

We note here that we have also checked the time evolution in the case where the $A_a^\eta$ and $\partial_\tau A_a^\eta$ were randomly sampled from the interval $[-\tau_0^{-1},\tau_0^{-1}]$ and found that the behaviour is as expected. $P_L/\epsilon$ starts at small values, either immediately in the quadratically growing phase or quickly transitioning to it. Since the interaction dominated phase starts early, $P_L/\epsilon$ transitions to oscillatory behaviour when it is still quite small. The late time phenomenology in this case is less rich: it always proceeds in a similar way to the solutions in Fig.~\ref{fig:pressure_early} with the lowest late time $P_L/\epsilon$.

\begin{figure}
    \centering
    \includegraphics[width=\linewidth]{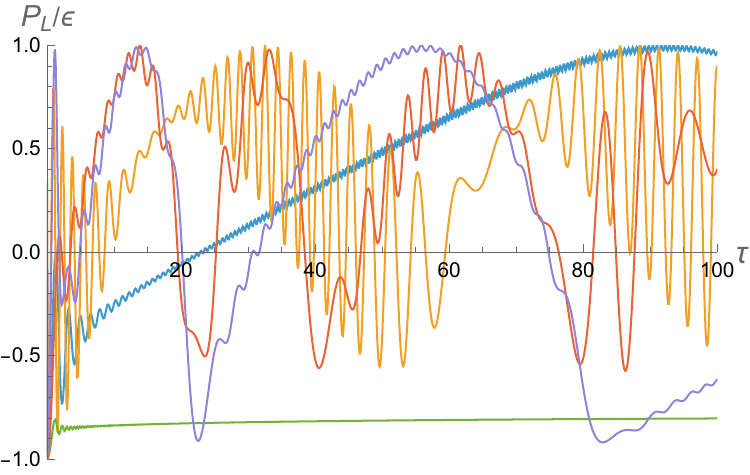}
    \caption{Late time evolution of the pressure anisotropy quantified by $P_L/\epsilon$ from numerical simulations of the dynamics prescribed by Eqs.~\eqref{eq:evolution_i} and~\eqref{eq:evolution_eta} for a representative subset of five of the set of random initial conditions from Fig.~\ref{fig:pressure_early}.}
    \label{fig:pressure_late}
\end{figure}

Fig.~\ref{fig:pressure_late} shows the evolution of the pressure ratio $P_L/\epsilon$ at late times, for five of the 100 initial conditions, which are representative of the range of observed behaviour. In this regime, scaling time by $\tau_{\rm int}$ is no longer sensible, as it can grow faster than $\tau$, and the behaviour of $P_L/\epsilon$ at different times but the same $\tau/\tau_{\rm int}$ is in no clear relation. 

Typically, the pressure ratio oscillates in a complex superposition of low and high frequencies. There is no indication of convergence to a late time fixed point. The purple curve seems to relax to a constant, however the slow growth in combination with the observed behaviour in the other curves suggests that a very slow oscillation is present.

Altogether, the diagnosis based on the pressure anisotropy paints the following picture. The purely expansion driven case at early times has a fixed point at $P_L/\epsilon=-1$ that is approached as $\tau^{-1}$. On the timescale $\tau_{\rm trans}$, a repulsive behaviour sets in, where $P_L/\epsilon$ grows quadratically. In this regime, the underlying dynamics is still dominated by expansion, but contributions to the pressure anisotropy are dominated by interaction. For $\tau\gtrsim\tau_{\rm int}$, interaction dominated dynamics show a complex oscillation pattern with no apparent universality.

However, the pressure anisotropy far from exhausts the phenomenology of this system. Even though no late time attraction is apparent in $P_L/\epsilon$, it may still manifest in other directions of state space. We know, for example, that the early time power law decay and growth are in fact separate phenomena. The decay in transverse electric fields persists even after transverse magnetic fields start dominating over them, meaning that this direction of state space is still contracting. In the regime of interaction domination, this flattened direction may be rotated in state space without showing actual divergence of neighbouring curves, meaning that a dimensional reduction persists. In order to elucidate this behaviour, it is necessary to perform a time resolved principal component analysis of sets of neighbouring trajectories in state space.

\section{Dimensional reduction}\label{sec:dimensional_reduction}

In higher dimensional state space, attractor behaviour can be diagnosed by  performing principal component analyses on a set of evolving trajectories in that state space at different points in time. The decrease of one of the principal component variances relative to the others will point to the loss of sensitivity to the associated state space direction. This diagnostic hinges on the set of trajectories deforming to a shape that is convex, ideally elliptic, so that bendings of its shape do not introduce additional variance. This can be achieved by starting with a set of initial conditions that has a radius in state space small enough that the nonlinear parts of the evolution equations are sufficiently similar for all trajectories. The state space that we consider are all of the 18 fields $E_a^i$, $E_{L,a}$, $B_a^i$, $B_{L,a}$ (see Eqs.~\eqref{eq:Edef} and~\eqref{eq:Bdef}). The imposed constraints~\eqref{eq:evolution_t} and~\eqref{eq:gaugeB} will cause the six variances in unphysical directions to be small, and we only consider the twelve largest variances. We choose these fields as the state space basis to work with because they are the physically relevant degrees of freedom that enter also the components of the energy-momentum tensor.

In order to avoid any accidental initial bias towards a direction in state space, we pick spherical ensembles of initial states. Details of their construction are found in App.~\ref{app:spherical_initial_ensembles}. We first consider an example centered around an initial condition of particularly simple form for ease of interpretation. Then we consider two more complex ensembles, centered each around an initial condition whose late time evolution of $P_L/\epsilon$ displayed either highly oscillatory behaviour or overall small values.

\begin{figure}
    \centering
    \includegraphics[width=\linewidth]{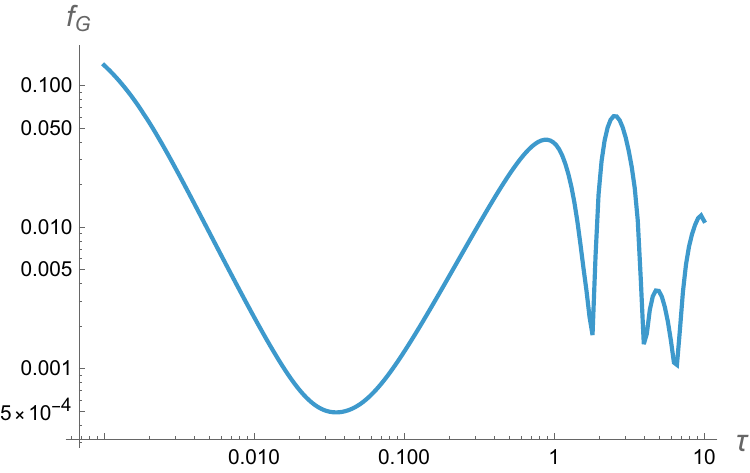}
    \caption{Time evolution of the upper estimate $f_G$ of the fraction of variance in the gauge subspace for the ensemble centered around the simple initial condition.}
    \label{fig:fG}
\end{figure}

The simple initial condition is given by the initial values $B_{L,3}=B_{T,1}^y=B_{T,2}^y=B_{T,2}^x=1$, while all other components are zero. Before discussing how the corresponding ensemble evolves in state space, we check one complication of the gauge choice:~\eqref{eq:gaugeB} is employed at initial time, but over time the gauge degrees of freedom may again be populated dynamically. We check the relevance of this effect to the ensemble variances via an upper estimate $f_G$ of the fraction of variance that is in the gauge degrees of freedom. Its construction is laid out in App.~\ref{app:gauge_subspace}. As can be seen in Fig.~\ref{fig:fG}, $f_G$ remains small throughout the full time evolution. It cannot be expected to fully vanish, because the gauge degrees of freedom will take slightly different forms for the different trajectories in the ensemble. The initial value gives an idea of the accuracy, as the actual population of gauge degrees of freedom is zero there. We conclude that we can consider this additional variance to be negligible.

\begin{figure}
    \centering
    \includegraphics[width=\linewidth]{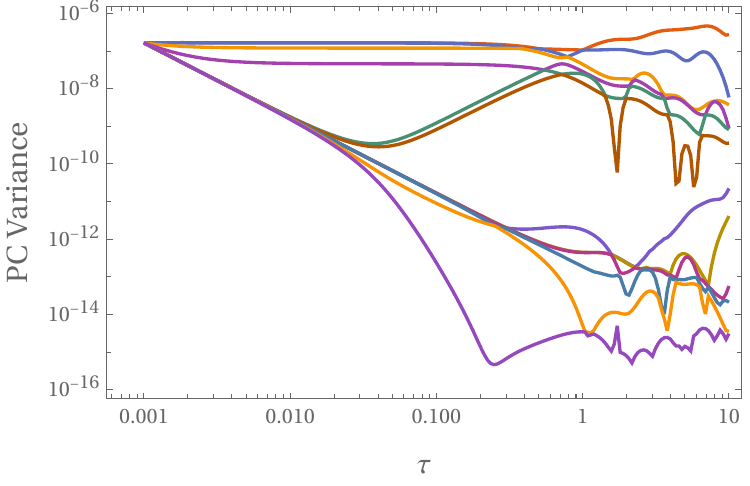}
    \caption{Evolution of principal component variances for the ensemble centered around the simple initial condition.}
    \label{fig:PCA_simple}
\end{figure}

The evolution of principal component variances for the ensemble centered around the simple initial condition is shown in Fig.~\ref{fig:PCA_simple}. At early times, the evolution splits into constant variances and those that decay $\propto\tau^{-2}$. The latter later split again into those that keep following the same decay, those that transition to an increase $\propto\tau^2$ and one that decays even faster. Except for the latter, this behaviour is the same as that in $P_L/\epsilon$, which was obtained from the squares of the field components. This is expected: if the whole ensemble of trajectories follows the same trend, then the variance of the ensemble will behave as the square of that trend. The faster decay in one variance comes from correlations in the $\propto\tau^2$ growth causing a shearing deformation in state space, which creates a narrow direction. The logarithmic terms cause small deviations from the expected early time behaviour, resulting in a degeneracy in the constant variances. At late times, interactions cause oscillating behaviour in the variances but do not significantly change the hierarchy of variances. Thus, in this regime information may be shuffled in state space, but the importance of the directions in state space given by the principal component eigenvectors is largely unchanged.

\begin{figure}
    \centering
    \includegraphics[width=\linewidth]{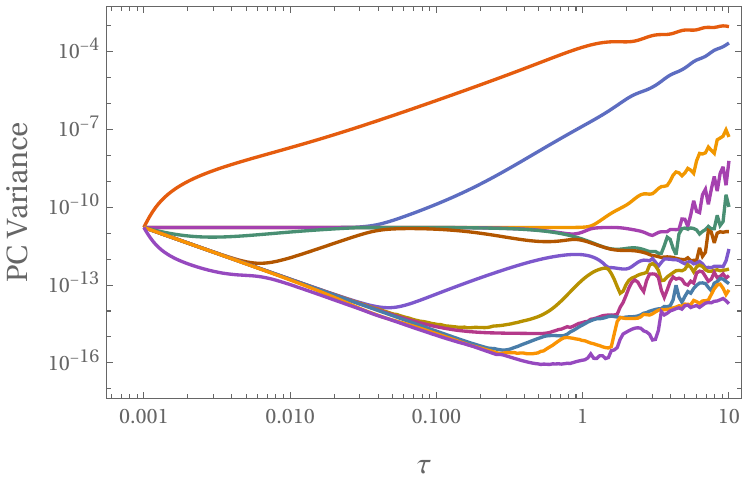}\\
    \includegraphics[width=\linewidth]{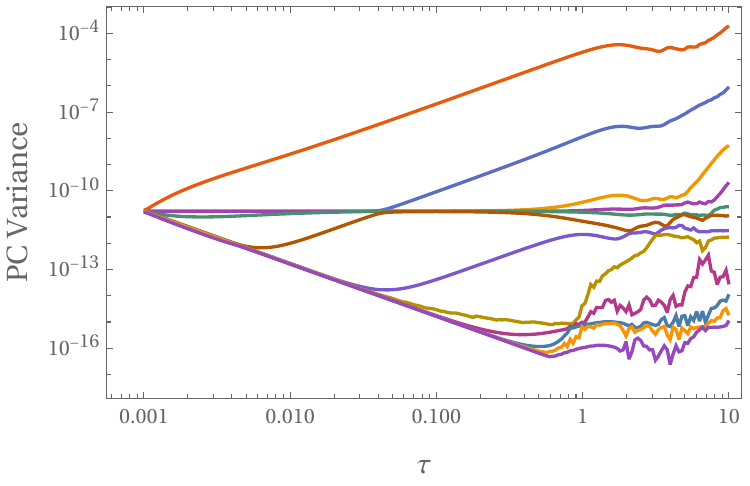}
    \caption{Evolution of principal component variances for the two ensembles centered around (top) the highly oscillating solution and (bottom) the solution with low late time $P_L/\epsilon$.}
    \label{fig:PCA_complex}
\end{figure}

Fig.~\ref{fig:PCA_complex} shows the evolution of principal component variances in the two more complex ensembles. They show quick changes at the earliest times, but otherwise seem qualitatively in agreement with the simple case. Again, decreasing and growing components split at early times, while the late time interaction dominated part causes mild oscillations in one case and strong ones in the other, which however do not change the hierarchy much.

\begin{figure}
    \centering
    \includegraphics[width=\linewidth]{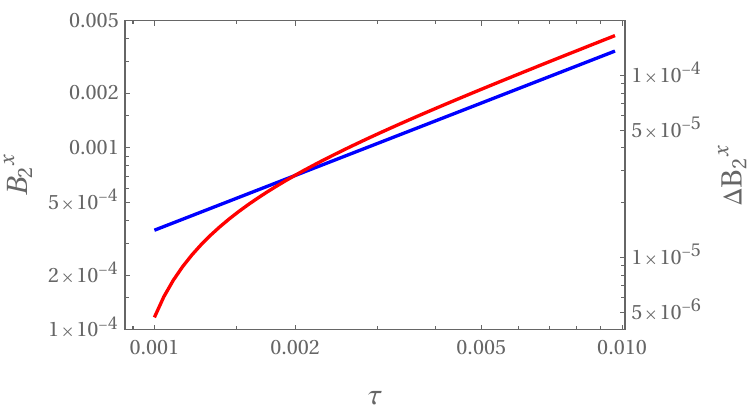}\\
    \includegraphics[width=\linewidth]{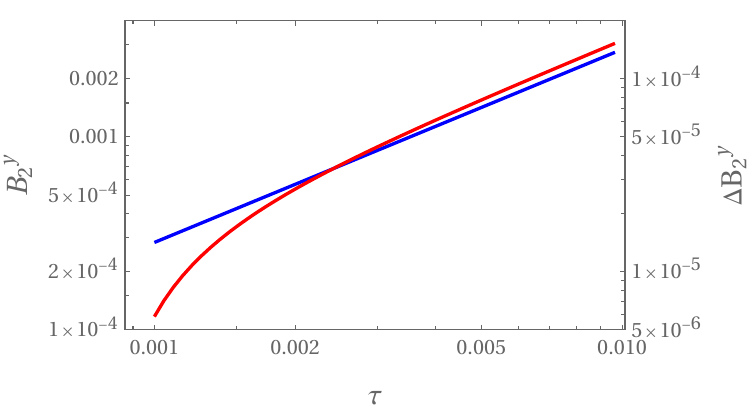}
    \caption{Time evolution of individual transverse magnetic fields from the (top) oscillating and (bottom) low $P_L/\epsilon$ ensembles used in PCA. Blue lines show the central trajectory, red lines show the biggest difference of any other ensemble trajectory to the central one.}
    \label{fig:tevo}
\end{figure}

While qualitatively the picture is always similar, the details of how the evolution of variances plays out shows some initial state dependence. Both of the examples feature a variance in one principal component that is growing quickly already at early times. The growth seems compatible with a $\tau^2$ power law and indeed, as shown in Fig.~\ref{fig:tevo}, a closer look at the evolution of the fields with the biggest components of the corresponding eigenvector reveals that in these cases one of the transverse magnetic fields grows quadratically in time already at initial time, and differences to neighbouring trajectories follow suit. In the oscillating case, multiple fields exhibit a correlated quadratic growth, which again causes a shearing deformation in state space and an emergent decreasing variance.

Depending on what features of the system one is interested in, the quick growth of one principal component may actually be beneficial, if said features are already sufficiently described by the behaviour of just this component. Of course, for the system considered here, this growth occurs only for a certain subset of initial conditions. It may, however, have causes other than an early onset of the quadratic growth. Another cause may be a configuration where some fields are much smaller than others, and their coupling causes them to start the oscillating behaviour already at initial time. Neighbouring trajectories will then start with different phases of the oscillation and start diverging. A very quick increase in variance can result from growth $\propto c+\ln\left(\tau/\tau_0\right)$ as seen in Eqs.~\eqref{eq:ana_BT} and~\eqref{eq:ana_BL}. For some of the differences between neighbouring trajectories, the constants cancel, leaving only the quickly growing $\propto \ln\left(\tau/\tau_0\right)$-term.

\section{Very late time growth}\label{sec:late_time}

In the Glasma, chaotic behaviour has been observed at late times~\cite{Romatschke:2005pm,Pooja:2026hjg}, which is associated with an exponential growth of differences between neighbouring trajectories. Though this exponential behaviour may be modified in an expanding geometry, as Ref.~\cite{Pooja:2026hjg} has observed growth with the exponential of the square root of time. In Sec.~\ref{sec:dimensional_reduction}, we found only power law growth, however due to numerical constraints the discussion was restricted to early times. We now want to take a closer look at late times, using the same simple initial condition for the center of our ensemble as before. To avoid numerical troubles, we tune the setup to the late time behaviour by starting with an increased scale of initial values for the fields -- i.e. larger $Q_s$ -- at $\tau_0=1$.

\begin{figure}
    \centering
    \includegraphics[width=\linewidth]{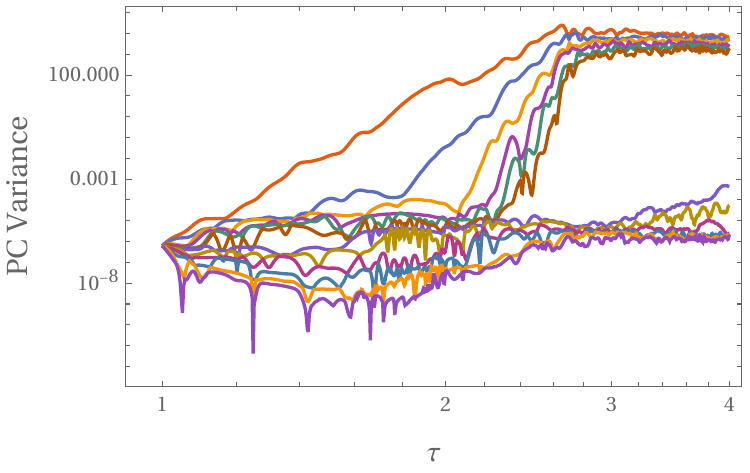}
    \includegraphics[width=\linewidth]{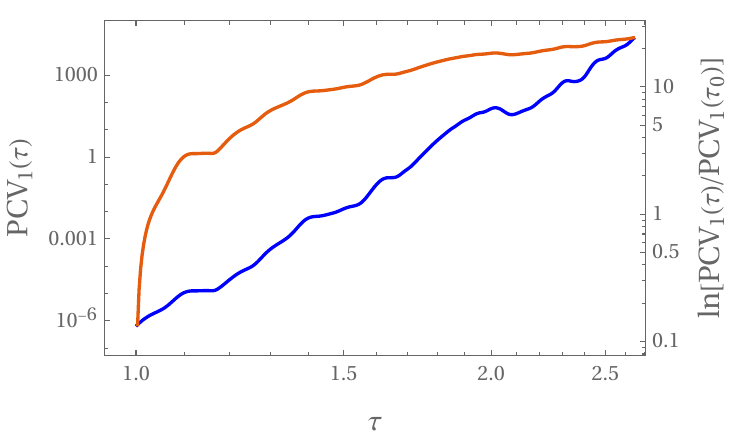}
    \caption{Time evolution of principal component variances in the late time case, facilitated by large initial values. The top plot shows all 12 physical variances, while the bottom plot shows the largest variance $\mathrm{PCV}_1$ as well as its logarithm $\ln\left(\frac{\mathrm{PCV_1(\tau)}}{\mathrm{PCV_1(\tau_0)}}\right)$.}
    \label{fig:PCA_late}
\end{figure}

The evolution of variances for the case $Q_s=100$ is shown in Fig.~\ref{fig:PCA_late}. In the cases of $Q_s=10$ and $Q_s=1000$, the resulting evolution is qualitatively very similar, so it is not shown here. Evidently, there is a strong growth in one of the variances. Other variances start growing later at an even faster rate, though this may be triggered by coupling to the other growing fields. Curiously, all growing variances saturate after some time to values on the same magnitude.

On first glance, it is unclear whether to classify the growth of the largest principal component variance $\mathrm{PCV}_1$ as exponential- or power law-like, since oscillatory behaviour is superimposed on the growth and the curves look compatible with a straight line in both log-log and log-linear scales. However, the distinction becomes obvious once the evolution of $\ln\left(\frac{\mathrm{PCV_1(\tau)}}{\mathrm{PCV_1(\tau_0)}}\right)$ is plotted in log-log scale, as in the bottom plot of Fig.~\ref{fig:PCA_late}. If the time dependence were as $\mathrm{PCV_1(\tau)}\propto a\exp[bt^\kappa]$, then this plot should be a straight line, but the plot is far from straight. We conclude that the time dependence is a power law $\propto at^\kappa$.

However, when comparing results for different $Q_s$, it seems that the exponent of this power law depends on $Q_s$. As seen in Table~\ref{tab:Qs_scaling}, both the exponent $\kappa$ and the time $\Delta\tau_{\rm sat}$ that the variances saturate seem to scale with powers of $Q_s$. In the Glasma, this would be unexpected, as $Q_s$ is the only dimensionful scale in the system and cannot be multiplied by another scale to obtain a dimensionless parameter. In the present case, however, $\tau_0$ provides another scale, though the dependence remains peculiar. As this work focuses on attractor behaviour, the explanation of these $Q_s$-dependences is left for future endeavors.  

\begin{table}[]
    \centering
    \begin{tabular}{c|cccc}
        $Q_s$ & $\kappa$& $\kappa Q_s^{-1/3}$ &$\Delta\tau_{\rm plat}$&$\Delta\tau_{\rm plat}Q_s^{1/2}$ \\
        \hline
        10 & $12$ & $5.4$ & $6.1$ & $19$ \\
        100 & $26$ & $5.5$ & $1.7$ & $17$ \\
        1000 & $58$ & $5.8$ & $0.53$ & $17$ 
    \end{tabular}
    \caption{$Q_s$-dependence and inferred scaling of the power law exponent $\kappa$ and the time $\Delta\tau_{\rm sat}$ of variance saturation in the late time evolution of the largest principal component.}
    \label{tab:Qs_scaling}
\end{table}

\section{Conclusions and Outlook}\label{sec:conclusions}

We have derived the evolution equations of classical Yang-Mills theory in a boost-invariant and transversely homogeneous system. Some of the equations become constraints for the fields that are conserved by the evolution, reducing the number of degrees of freedom of this system. The pressure anisotropy of this system exhibits an early time attractor in the form of power law decay to $P_L/\epsilon=-1$. When interaction driven field components become important, this behaviour transitions to a power law growth. Even later, interactions dominate also the time evolution, and $P_L/\epsilon$ oscillates in a superposition of small and large frequencies.

This indicates that there are three stages: first, expansion drives memory loss. Then, expansion dominated evolution with interaction dominated degrees of freedom partly restores memory. Finally, interaction dominated dynamics shuffle information in state space with no clear memory loss or restoration. Thus, in this system, expansion is the only memory loss mechanism. The partial restoration of memory makes it an interesting example case.

The state space picture confirms qualitatively the behaviour in these three stages. There is an initial stage of dimensional reduction, then a partial restoration of state space directions, and later an oscillatory behaviour that keeps the hierarchy of state space directions largely intact. State space directions that lost importance in the first stage and did not regain it in the second remain irrelevant at late times. However, the microscopic reasons for this emerging hierarchy may differ from the simple early time explanation and depends strongly on the initial state. In particular, this happens when for some field components either the timescale hierarchy of expansion dominated and interaction dominated dynamics does not hold, or finite-$\tau_0$-effects cannot be neglected.

In MIS hydrodynamics, holographic descriptions and most kinetic theories, the late time dynamics of Bjorken flow are always characterized by attractive behaviour towards equilibrium. Non-fluctuating classical Yang-Mills theory is the first example of an interacting dynamical prescription that exhibits only expansion driven memory loss. This is in contrast to the Glasma, which through interactions exhibits attraction to a universal self-similar time evolution~\cite{Berges:2013eia,Berges:2013fga}. The new finding compliments ongoing endeavors to generalize the concept of the attractor to other expansion geometries like Gubser flow~\cite{Denicol:2018pak,Behtash:2019qtk,Dash:2020zqx} or FLRW~\cite{Buza:2024jxe}, or to a completely different system with the hydrodynamic attractor in cold atoms~\cite{Fujii:2024yce,Mazeliauskas:2025jyi,Heller:2025yxm}. For future studies aiming at general characterizations of attractor behaviour or their exploitation for effective dynamical descriptions in the spirit of Ref.~\cite{DuPlessis:2026qjy}, the system studied here will be an important test case.

\begin{acknowledgements}
    The author thanks Dana Avramescu, Kirill Boguslavski, Pooja, Tuomas Lappi, Micha\l\ Heller and Sören Schlichting for helpful discussions. This project has received funding from the European Research Council (ERC) under the European Union’s Horizon 2020 research and innovation programme (grant number: 101089093 / project acronym: High-TheQ). Views and opinions expressed are however those of the authors only and do not necessarily reflect those of the European Union or the European Research Council. Neither the European Union nor the granting authority can be held responsible for them.
\end{acknowledgements}

\newpage

\appendix

\section{Derivation of Evolution Equations}

The symmetry assumptions made in this work largely simplify the dynamics of the system we consider. To recap, we assume boost invariance and transverse homogeneity. In Milne coordinates $(\tau,x,y,\eta)$ these assumptions can be expressed as $A^\mu(\tau,x,y,\eta)=A^\mu(\tau)$. These coordinates are further characterized by the metric $g_{\mu\nu}=\mathrm{diag}(1,-1,-1,-\tau^2)$ and
\begin{align}
    \Gamma^\tau_{\eta\eta}=\tau\,, \quad \Gamma^\eta_{\tau\eta}=\Gamma^{\eta}_{\eta\tau}=\tau^{-1}\,,
\end{align}
while all other Christoffel symbols vanish. We further adopt the gauge choice $A^\tau_a=0$. In this setup, the field strength tensor 
\begin{align}
    F^{\mu\nu}&=\nabla^\mu A^\nu-\nabla^\nu A^\mu-ig[A^\mu,A^\nu]\\
    F_a^{\mu\nu}&=\partial^\mu A^\nu + \Gamma^{\mu \ \ \nu}_{\ \ \lambda} A^\lambda -\partial^\nu A^\mu - \Gamma^{\nu \ \ \mu}_{\ \ \lambda} A^\lambda +gf_a^{bc}A_b^\mu A_c^\nu
\end{align}
as stated in Eqs.~\eqref{eq:Ftaui}--\eqref{eq:Fuv} simplifies to
\begin{align}
    F_a^{\tau i}&=\partial_\tau A_a^i-\underbrace{{\partial^iA_a^\tau}}_{=0}+\underbrace{(\Gamma^{\tau \ \ i}_{\ \ \lambda}-\Gamma^{i \ \ \tau}_{\ \ \lambda})}_{=0}A_a^\lambda+gf_a^{bc}\underbrace{A_b^\tau}_{=0} A_c^i\nonumber\\
    &=\partial_\tau A_a^i\,,\\
    F_a^{\tau \eta}&=\partial_\tau A_a^\eta-\underbrace{{\partial^\eta A_a^\tau}}_{=0}+(\Gamma^{\tau \ \ \eta}_{\ \ \lambda}-\Gamma^{\eta \ \ \tau}_{\ \ \lambda})A_a^\lambda+gf_a^{bc}\underbrace{A_b^\tau}_{=0} A_c^\eta\nonumber\\
    &=\partial_\tau A_a^\eta+2\tau^{-1}A_a^\eta\,,\\
    F_a^{uv}&=\underbrace{\partial^u A_a^v}_{=0}-\underbrace{{\partial^v A_a^u}}_{=0}+\underbrace{(\Gamma^{u \ \ v}_{\ \ \lambda}-\Gamma^{v \ \ u}_{\ \ \lambda})}_{=0}A_a^\lambda+gf_a^{bc}A_b^u A_c^v\nonumber\\
    &=gf_a^{bc}A_b^u A_c^v\,.
\end{align}
Time evolution of the field strength tensor is dictated by
\begin{align}
   0&= D_\mu F^{\mu\nu}\\
   0&=\partial_\mu F_a^{\mu\nu} + \Gamma^{\mu}_{\ \ \lambda\mu}F_a^{\lambda\nu}+\Gamma^{\nu}_{\ \ \lambda\mu}F_a^{\mu\lambda} +gf_a^{bc}A_{b,\mu} F^{\mu\nu}_c\,.
\end{align}
\begin{widetext}
    For our system, this simplifies to Eqs.~\eqref{eq:evolution_t}--\eqref{eq:evolution_eta} as stated below.
\begin{align}
    D_\mu F^{\mu\tau}_a&=\underbrace{\partial_\mu F^{\mu\tau}_a}_{=0}+\underbrace{\Gamma^\mu_{\ \ \lambda \mu}F_a^{\lambda \tau}}_{=0}+\underbrace{\Gamma^\tau_{\ \ \lambda \mu}F_a^{\mu\lambda}}_{=0}+g f_a^{bc}\left(
    A_{b,i}F_c^{i\tau}+A_{b,\eta}F^{\eta\tau}_a\right)\\
    &=gf_a^{bc}\left(A_b^i\partial_\tau A_c^i+\tau^2A^\eta_b\partial_\tau A^\eta_c\right)\\
    D_\mu F^{\mu i}&=\underbrace{\partial_\mu F^{\mu i}_a}_{=\partial_\tau^2A_a^i}+\underbrace{\Gamma^\mu_{\ \ \lambda \mu}F_a^{\lambda i}}_{=\tau^{-1}\partial_\tau A_a^i}+\underbrace{\Gamma^i_{\ \ \lambda \mu}F_a^{\mu\lambda}}_{=0}+g f_a^{bc}\left(
    \underbrace{A_{b,\tau}}_{=0}F_c^{\tau i}+A_{b,u}F^{ui}_c\right)\\
    &=\partial_\tau^2A_a^i+\tau^{-1}\partial_\tau A_a^i-g^2f_a^{bc}f_c^{de}(A^{\bar{i}}_bA^{\bar{i}}_d+\tau^2A_b^\eta A_d^\eta)A_e^i\\
    D_\mu F^{\mu \eta}&=\underbrace{\partial_\mu F^{\mu \eta}_a}_{=\partial_\tau F_a^{\tau\eta}}+\underbrace{\Gamma^\mu_{\ \ \lambda \mu}F_a^{\lambda i}}_{=\tau^{-1}F_a^{\tau\eta}}+\underbrace{\Gamma^\eta_{\ \ \lambda \mu}F_a^{\mu\lambda}}_{=0}+g f_a^{bc}\left(
    \underbrace{A_{b,\tau}}_{=0}F_c^{\tau \eta}+A_{b,u}F^{u\eta}_c\right)\\
    &=\partial_\tau^2A_a^\eta+2\tau^{-1}\partial_\tau A_a^\eta-2\tau^{-1} A_a^\eta+\tau^{-1}\partial_\tau A_a^\eta+2\tau^{-1} A_a^\eta -g^2 f_a^{bc}f_c^{de}A^i_bA^i_dA_e^\eta\\
    &=\partial_\tau^2A_a^\eta+3\tau^{-1}\partial_\tau A_a^\eta -g^2 f_a^{bc}f_c^{de}A^i_bA^i_dA_e^\eta
\end{align}

\end{widetext}

\section{Inverting commutator relation}\label{app:inverting}

Section~\ref{sec:setup} mentions the necessity to invert the equations determining the field strength tensor $F_a^{\mu\nu}$ in terms of the gauge fields $A_a^\mu$. This is necessary for translating specific points in the chosen state space into the dynamical quantities entering the evolution equations. The hardest part of this inversion is the following equation:
\begin{align}
    F_a^{uv}&=gf_a^{bc}A_b^uA_c^v\,.\label{eq:Fuv_app}
\end{align}

In fact, it is not possible to solve for the $A_a^u$ in general: if $B\in \mathfrak{su}(N)$ commutes with all three $A^u$, then adding it to them
\begin{align}
    gf_{a}^{bc}A_b^u(A_c^v+B_c)=gf_{a}^{bc}A_b^uA_c^v=F_a^{uv}
\end{align}
will result in the same $F_a^{uv}$, i.e. there is a flat direction in the map $\{A_a^u\}\mapsto \{F_a^{uv}\}$. Since both sides have the same number of unknowns, $3(N_c^2-1)$, we have to conclude that some combinations of values for the $F_a^{uv}$ cannot be written in the form of Eqs.~\eqref{eq:Fuv} and are therefore unphysical under our chosen symmetry constraints. 

However, in the special case $N_c=2$, Eqs.~\eqref{eq:Fuv} can be algebraically solved for $A_a^u$. We can make use of the fact that $f_a^{bc}f_c^{de}=\delta_a^d\delta^{be}-\delta_a^e\delta^{bd}$ to arrive at
\begin{align}
    A_a^u&=\frac{1}{\sqrt{N}}f_a^{bc}F_b^{u,(u+1)\%3}F_c^{(u-1)\%3,u}\,,\\
    N&=gf^{abc}F_a^{xy}F_b^{y\eta}F_c^{\eta x}\,.
\end{align}
We immediately see that in order to obtain real-valued $A_a^u$, we need $N>0$. Again, some higher-dimensional quadrants in the space of the $F_a^{\mu\nu}$ seem to be unphysical. This is related to the fact that a global shift $A_a^u\to -A_a^u$ leaves the $F_a^{uv}$ invariant as they are quadratic in the $A_a^u$. This ambiguity is reflected in the freedom of choice for the sign of the square root.

\section{Spherical Initial Ensembles}\label{app:spherical_initial_ensembles}

In order to meaningfully compare the evolution of principal component variances, the initial ensemble should be spherical in the sense that the covariance matrix of this ensemble in state space is isotropic. Because of the large number of directions in the state space we consider, random sampling would require a high amount of initial states in order to converge to a spherical shape, as demonstrated in Table~\ref{tab:sample_PCA}. Instead, we use a symmetric set of points on a sphere centered around a chosen initial condition, which we construct as follows.

\begin{table}[]
    \centering
    \begin{tabular}{c|cc}
        N & Gauss & Sphere \\
         \hline
        $10^3$ & $35\% $ & $31\% $ \\
        $10^4$ & $12.2\% $ & $11.0\% $ \\
        $10^5$ & $3.8\% $ & $3.7\% $ 
    \end{tabular}
    \caption{Deviations from isotropy of random isotropic samples in 12-dimensional space. For the column ``Gauss'', points were sampled from an isotropic Gaussian. for the column ``Sphere'', these points were projected onto the unit sphere by dividing by their distance from the origin. The values show the relative difference between the largest and smallest variance in different directions of space, as obtained by principal component analysis. They are compatible with a scaling with $N^{-1/2}$.}
    \label{tab:sample_PCA}
\end{table}

First, we need to find an orthonormal basis on the tangent space to the physical subspace that satisfies the constraints~\eqref{eq:evolution_t} and~\eqref{eq:gaugeB} at the position of the chosen center of our ensemble. We can not simply construct this basis in the coordinate map of this subspace because the coordinate mapping is not an isometry, so different points at fixed distance in the subspace coordinates will not have the same distance in ambient space. To do this, we add small perturbations to the subspace coordinates of the center point $X$ and find the resulting difference in the ambient space coordinates, $\delta X_i=X_i'-X$, for all directions in subspace, $i=1,...,12$. The tangent space basis $e_i$ is then found by orthonormalization of the set of $\delta X_i$.

We then construct our ensemble of initial states as a set of points on a sphere of small radius around our chosen center initial condition. In order to result in an isotropic covariance matrix, the point set must be highly symmetric. These points can be obtained as the midpoints of boundary structures of a fitting hypercube (vertices, edges, faces, ...). The number of $d$-dimensional such boundary structures for an $n$-dimensional hypercube is $2^{n-d}\binom{n}{d}$. For the unit sphere, the points can be found by finding all ways of distributing $d$ values of $\pm 1/\sqrt{d}$ among the $n$ vector components, with the other ones being zero. In our case, with 12 degrees of freedom in the initial state, the simplest choice is $d=11$ with $24$ points, one in the direction of each unit vector and one in the opposite direction. In order to avoid the accurate representation of a state space direction hinging on the numerical accuracy of a single trajectory, we pick the next to simplest case of $d=10$ with $264$ points. To this we add the initial condition in the center of the sphere. This construction ensures that all directions in state space start with the same variance and the sensitivity to all directions is the same.  

\section{Finding the gauge subspace}\label{app:gauge_subspace}

Even though gauge degrees of freedom are eliminated at initial time, because of the fact that no closed form can be found for the residual gauge fixing on top of $A^\tau=0$, the evolution equations do not preserve it and the gauge degrees of freedom may be populated again later in time. In order to probe to which extent this happens, we first need to find the gauge subspace at later times. For any state $X_a=(E_{T,a}^x,E_{T,a}^y,E_{L,a},B_{T,a}^x,B_{T,a}^y,B_{L,a})$ of the system, we can find three vectors of the local tangent gauge subspace as
\begin{align}
    g_a^{(b)}=f_a^{bc}X_c\,.
\end{align}
We can then orthonormalize and collect them in a $18\times 3$-matrix $G$ that can be used for projection onto the gauge subspace. The covariance matrix $\Sigma$ of an ensemble in state space may be projected as $G\Sigma G^T$. From this, we define the fraction of its variance that lies in the gauge subspace as 
\begin{align}
    f_G=\frac{\mathrm{tr}\,G\Sigma G^T}{\mathrm{tr}\,\Sigma}\,.
\end{align}
Since the projection matrix is constructed for the central trajectory, while the gauge tangent space may be oriented slightly differently for the other trajectories in the ensemble, this quantity is not zero even at initial time, as seen in Fig.~\ref{fig:fG}.

\bibliography{YM}

\end{document}